# 120-MM SUPERCONDUCTING QUADRUPOLE FOR INTERACTION REGIONS OF HADRON COLLIDERS*

A.V. Zlobin#, V.V. Kashikhin, N.V. Mokhov, I. Novitski, Fermilab, Batavia, IL 60510, U.S.A.

*Abstract*

Magnetic and mechanical designs of a $Nb_3Sn$ quadrupole magnet with 120-mm aperture suitable for interaction regions of hadron colliders are presented. The magnet is based on a two-layer shell-type coil and a cold iron yoke. Special spacers made of a low-Z material are implemented in the coil mid-planes to reduce the level of radiation heat deposition and radiation dose in the coil. The quadrupole mechanical structure is based on aluminum collars supported by an iron yoke and a stainless steel skin. Magnet parameters including maximum field gradient and field harmonics, $Nb_3Sn$ coil pre-stress and protection at the operating temperatures of 4.5 and 1.9 K are reported. The level and distribution of radiation heat deposition in the coil and other magnet components are discussed.

## INTRODUCTION

Low-beta quadrupoles used in interaction regions (IRs) of hadron colliders are exposed to a strong radiation produced by secondary particles coming from the interaction point. This radiation leads to a high level of heat deposition in the coils and total heat load in the magnet cold mass, and causes a high radiation dose in the magnet structural materials. Reliable IR magnet operation and acceptable life-time require an increased magnet operation margin, additional cryogenic power for triplet cooling and special structural materials with high radiation resistance. Taking into account the fact that radiation heat deposition in IR low-beta quadrupoles is localized in the magnet mid-planes, the above problems can be partially solved by using low-Z materials and minimizing the material volume in these areas. This approach has been previously used in the design of a separation dipole D1 proposed for the LHC IR upgrade [1]. In this paper, the abovementioned approach is applied to the IR quadrupoles suitable for the LHC luminosity upgrade [2]. For a consistent comparison with the LHC upgrade scenario described in [3], the aperture of $Nb_3Sn$ IR quadrupoles is chosen to be 120 mm, with the yoke OD of 550 mm, and the nominal field gradient of 130 T/m. The quadrupole design parameters are compared to that of the traditional quadrupole design without spacers.

## QUADRUPOLE DESIGN

The quadrupole design is based on a two-layer shell-type coil and a cold iron yoke. The magnetic design optimization was performed using ROXIE code with non-linear properties of the iron yoke. The main optimization objectives were highest field gradient and field quality.

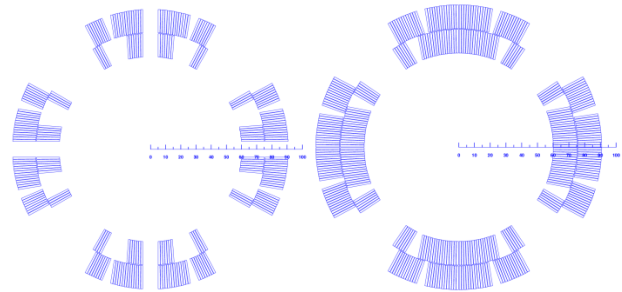

Figure 1: Cross-sections of the coils with (left) and without (right) mid-plane spacers.

Fig. 1 shows the cross-sections of coils with and without mid-plane spacers. The total spacer azimuthal thickness is 1 cm that allows the elimination of ~4 turns next to each coil mid-plane where the radiation heat and dose reach their maximum values. Both coils use a Rutherford-type cable with 40 strands of 0.7 mm in diameter. The bare cable width is 15.15 mm, and the small and large edges are 1.171 and 1.344 mm respectively. The cable insulation thickness is 0.1 mm.

The cross-section of a quadrupole based on coils with aluminium (Al) mid-plane spacers is shown in Fig. 2. The magnet mechanical structure consists of an octagonal Al collar, a two-piece iron yoke with large 105 mm holes for HeII heat exchanger, and a stainless steel skin. Two Al spacers control the gap between the two iron halves during magnet assembly and cool-down and house the yoke-skin alignment keys. The outer shape of the octagonal collar and the corresponding iron inner surface provide the alignment and surface matching at room and operation temperatures. Using the low-Z Al spacers and collar, placing the heat exchangers close to the coil mid-planes and removing the iron near the magnet mid-planes improve the coil cooling conditions and the radial heat transfer in the cold mass. However, participation of both these structural elements in the coil pre-stress and support is required.

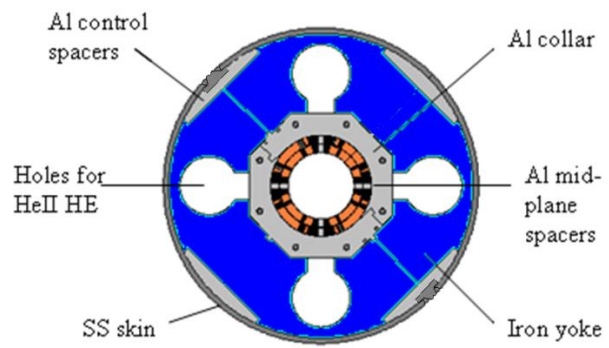

Figure 2: Quadrupole cold mass cross-section.

* Work supported by Fermi Research Alliance, LLC, under contract No. DE-AC02-07CH11359 with the U.S. Department of Energy
# zlobin@fnal.gov

## QUADRUPOLE PARAMETERS

The magnet quench parameters for the two coil designs calculated for the critical current density of $Nb_3Sn$ cable $J_c(12T, 4.2K)=2500$ A/mm$^2$ and Cu-to-nonCu ratio of 0.87 are summarized in Table 1. The smaller number of turns in the coil with mid-plane spacers (132 vs. 200 in the coil without spacers) causes the 18% reduction of its maximum gradient. Nevertheless, this design provides 23% margin with respect to the nominal operation gradient of 130 T/m even at 4.5 K. At 1.9 K the quadrupole with mid-plane spacers can provide 150 T/m gradient with almost 20% margin, sufficient to operate magnets at the luminosity of $10^{35}$ cm$^{-2}$s$^{-1}$ [4-5].

Table 1: Quadrupole Parameters at 4.5 and 1.9 K.

| Parameter | Operation temperature | Magnet design | |
|---|---|---|---|
| | | With spacer | Without spacer |
| Max gradient, T/m | | 159.6 | 189.9 |
| Quench current, kA | 4.5 K | 20.32 | 15.40 |
| Coil peak field, T | | 11.83 | 13.17 |
| Max gradient, T/m | | 174.5 | 208.0 |
| Quench current, kA | 1.9 K | 22.29 | 16.92 |
| Coil peak field, T | | 12.93 | 14.42 |

The optimized geometrical field harmonics at the reference radius of 40 mm and the gradient of 130 T/m for the two coil designs are presented in Table 2. Since the gap in the coil mid-planes is relatively large, the design with mid-plane spacers has a larger $b_{18}$ component which, however, becomes less than 1 unit at the reference radius of 39 mm.

Table 2: Geometrical Harmonics at $R_{ref}$=40 mm and G=130 T/m ($10^{-4}$).

| Harmonic | Magnet design | |
|---|---|---|
| | With spacer | Without spacer |
| $b_6$ | 0.0000 | 0.0000 |
| $b_{10}$ | 0.0039 | -0.0057 |
| $b_{14}$ | 0.0605 | -0.0278 |
| $b_{18}$ | -1.3996 | -0.4220 |

The yoke saturation effect on the load lines and the lowest order allowed harmonic $b_6$ for quadrupoles with and without coil mid-plane spacers is shown in Figs. 3-4.

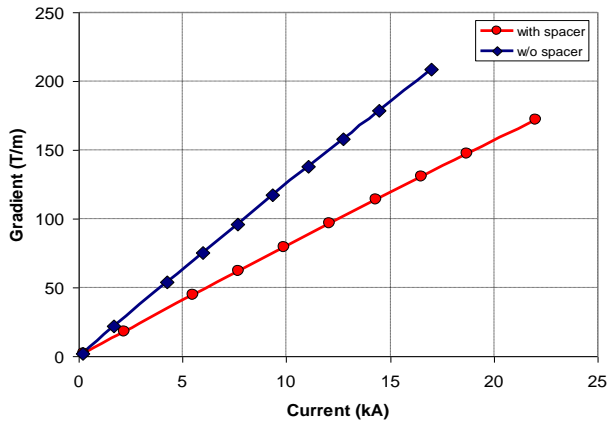

Figure 3: Quadrupole load lines.

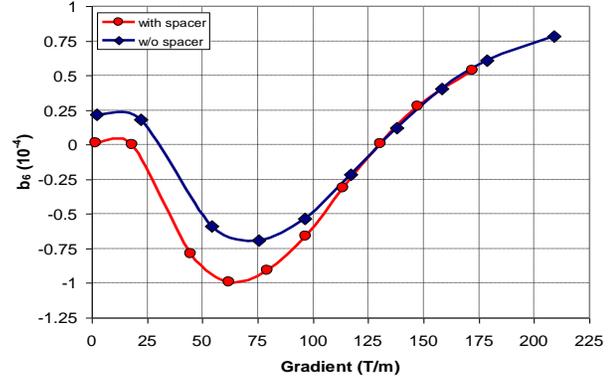

Figure 4: Yoke saturation effect on $b_6$.

Due to a relatively large distance between the coil and yoke, the iron saturation effect on the magnet load lines is small. The $b_6$ varies within +/- 0.5 units in the gradient range of 0-130 T/m that seems acceptable without an additional correction.

## COIL STRESS

Mechanical stresses in the quadrupole coil with mid-plane Al spacers were analyzed using ANSYS code with plastic coil material properties. The stress distributions in the coil at room temperature and at helium temperatures at zero current and at the field gradient of 130 T/m are shown in Fig. 5. The maximum room-temperature stress of 70 MPa in the coil after assembly increases after cool-down to ~110 MPa. This level of coil pre-stress allows the coil turns to be kept under compression at the nominal field gradient of 130 T/s. The maximum stress in the coil in this case does not exceed 130 MPa. To operate this magnet at the nominal field gradient of 150 T/m, the coil room temperature pre-stress needs to be increased to 90 MPa. The maximum coil stress at 150 T/m will not exceed 170 MPa which is safe for $Nb_3Sn$ magnets based on the recent studies performed using $Nb_3Sn$ quadrupole coils.

## QUENCH PROTECTION

The 120-mm $Nb_3Sn$ quadrupoles at the nominal gradient of 130 T/m have a nominal stored energy almost a factor of two higher than the present 70-mm IR quadrupoles (MQXB). That is why their protection needs to be carefully considered. A summary of the quench analysis for 120-mm $Nb_3Sn$ quadrupoles described above is reported in Table 3.

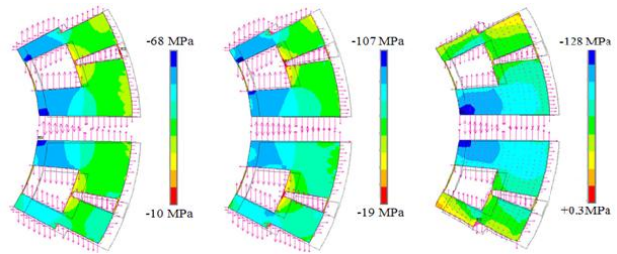

Figure 5: Coil stress at room temperature, at operation temperature and zero current, at 1.9 K and 130 T/m.

Table 3: Quench Protection Parameters.

| Parameter | Magnet design | |
|---|---|---|
| | With spacer | No spacer |
| Inductance (mH/m) | 4.1 | 9.0 |
| Stored energy @130T/m (kJ/m) | 558 | 488 |
| Operation current @130T/m (A) | 16430 | 10410 |
| Max. hotspot temperature (K) | 288 | 160 |
| Max. temperature under QH (K) | 132 | 117 |

Table 4: Radiation Dynamic Heat Load in Q1 (in Watts).

| Part | Magnet design | | |
|---|---|---|---|
| | With spacer | Without spacer | Reference* |
| Coil | 185 | 206 | 205 |
| Collar | 37 | 29 | 67 |
| Yoke | 79 | 72 | 42 |
| Total | 301 | 306 | 314 |

* Coil without mid-plane spacers with SS collar.

The analysis assumes a total heater delay of 30 ms. The quench heaters (QH) are placed on the coil outer layer and cover 60% of the turns in the coil with an Al spacer and 50% in the coil without a spacer. One can see that at 130 T/m all temperatures are below 300 K. At 150 T/m with the same heaters, the maximum hotspot temperature increases to 370 K and 180 K in coils with and without Al spacers respectively.

## RADIATION HEAT DEPOSITIONS

The radiation analysis was performed using MARS15 code [6] for the LHC inner triplet layout v.6.5. The four quadrupoles Q1, Q2A, Q2B, and Q3 have the same cross-sections and a 3.1 mm SS beam pipe. An additional 6.2 mm segmented SS liner was placed inside the Q1 beam pipe. The longitudinal distribution of maximum power density in the cross-section of quadrupoles with and without Al mid-plane spacers in the coil at the luminosity of $10^{35}$ cm$^{-2}$s$^{-1}$ is shown in Fig. 6. The peak power in the coil with Al spacer is reduced by almost a factor of two with respect to the traditional design and practically meets the design goal shown by the horizontal line.

The integrated power dissipation in the main components of Q1 for designs with Al collar, and with and without Al spacers is shown in Table 4 and compared with the reference quadrupole without mid-plane spacers and with SS collar [3, 5]. The dynamic heat load in the coil with Al spacers is 11% lower than in the traditional coils. The power dissipation in the Al collar is also almost a factor of two lower than in the SS collar. More transparent for particles coil and Al collar increase the heat deposition in the iron yoke by a factor of two making the radial heat distribution more uniform. The total dynamic heat loads in the case with Al spacers and collar is 4% lower than in the traditional quadrupole design.

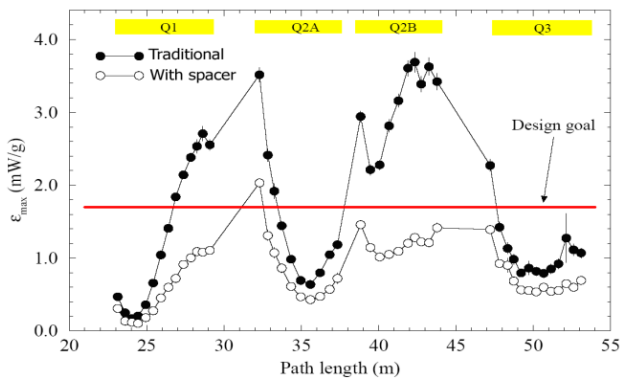

Figure 6: The peak power density in the coil inner layer of the quadrupole with and without Al spacers.

## SUMMARY

A 120-mm Nb$_3$Sn quadrupole magnet with Al mid-plane spacers and Al collar was analyzed and compared to the traditional magnet designs. It was found that the design with the Al mid-plane spacers can operate at the nominal gradient up to 150 T/m with operation margin ~20%. The mechanical analysis, performed for various operating conditions, demonstrates that the maximum stresses in the coil can be limited to a safe value of 170 MPa at the gradient of 150 T/m. The quench protection analysis shows that it is possible to operate the described Nb$_3$Sn quadrupoles with Al spacers at field gradients up to 150 T/m within the temperature limit of 370 K with quench heaters placed on the coil outer layer covering 60% of total coil turns.

The radiation and thermal analysis of IR quadrupole magnet with mid-plane spacers was performed and compared with the traditional quadrupole design. It was found that the Al mid-plane spacers reduce the peak power density and corresponding absorbed dose in the coil by almost a factor of two, which has a noticeable impact on the coil temperature margin and magnet lifetime. In addition, making the coil mid-plane more transparent to hadronic and electromagnetic showers has an extra benefit of redistributing the radiation heat deposition in the radial direction, improving the radial heat transfer in the quadrupole cold mass. The reduction of the total dynamic heat load on the cryogenic system is about 5%.